\documentclass[preprint2]{aastex631} 

\usepackage{natbib,amsmath}
\usepackage{comment}

\usepackage{tikz}
\usepackage{graphicx}
\usetikzlibrary{shapes.geometric, arrows,calc}
\newcommand{\lya}{Ly$\alpha$\relax}

\newcommand{\z}{\ensuremath{z}}

\newcommand{\autoslit}{{\tt autoslit}}
\newcommand{\tilsotua}{{\tt tilsotua}}
\newcommand{\slitastrometry}{{\tt slit\textunderscore astrometry}}

\shorttitle{Tilsotua for LRIS Multislit Archive}
\shortauthors{Sullivan et al.}

\begin{document}

\title{Tilsotua: Reproducing Slitmask Sky Positions for use with the LRIS Multislit Archive}

\author[0000-0002-9372-132X]{Jessica Sullivan}
\email{jsulli27@nd.edu}
\affiliation{Department of Physics and Astronomy, University of Notre Dame, Notre Dame, IN 46556}

\author[0000-0002-2591-3792]{J. Christopher Howk}
\affiliation{Department of Physics and Astronomy, University of Notre Dame, Notre Dame, IN 46556}

\author[0000-0001-9158-0829]{Nicolas Lehner}
\affiliation{Department of Physics and Astronomy, University of Notre Dame, Notre Dame, IN 46556}

\author[0000-0002-7893-1054]{John M. O'Meara}
\affiliation{W. M. Keck Observatory, Waimea, HI 96743}

\author[0000-0003-3801-1496]{Sunil Simha}
\affiliation{Department of Astronomy and Astrophysics, University of Chicago, Chicago, IL 60637}
\affiliation{Center for Exploration and Research in Astronomy, Northwestern University, Evanston, Chicago, IL 60201}
\affiliation{Department of Astronomy and Astrophysics, University of California Santa Cruz, Santa Cruz, CA 95064}
\keywords{Spectroscopy, Astronomy software, Software documentation, Publicly available software}
\begin{abstract}

We present \tilsotua{}, a code that calculates sky positions of the slits from Low-Resolution Imaging Spectrograph (LRIS) slitmasks used in multislit observations. Raw data for the Keck/LRIS spectrograph does not include information about the sky positions of the slitmask targets, making it difficult to use beyond the scope of the original programs. \tilsotua\ translates slit coordinates from the mask design files in the mask milling machine frame to sky coordinates. \tilsotua\ also shifts the original input astrometry to modern frames using the objects targeted by mask alignment boxes. This can be applied to the archived mask design files at the Lick Observatory Archive. We demonstrate that the final reconstructed slit positions are accurate to $0\farcs14$ (RMS) across the set of archived masks based on a comparison of our calculated mask alignment box coordinates with Gaia astrometric positions of the likely alignment objects. We make available the \tilsotua\ code, archived mask files, and slit positions of the historical masks for the community to maximize the science from Keck/LRIS data in the Keck Observatory Archive.
\end{abstract}

\section{Introduction}
\label{sec:intro}

The Low-Resolution Imaging Spectrograph (LRIS) on the Keck I telescope has been a world-leading optical spectrograph for nearly 30 years \citep[with design features summarized in][]{oke1995,mccarthy1998,steidel2004}, probing astrophysical objects from the present day to the epoch of reionization. LRIS's design \citep[described in][]{oke1995} characteristics have made it a powerful instrument for a wide range of science goals. LRIS has studied open cluster white dwarfs \citep{Kalirai07}, chemical abundances in Milky Way dwarf spheroidals \citep{bosler07,  shetrone10} and M31 globular clusters \citep{reitzel04}, high--z galaxy populations \citep{cohen00, reddy06,steidel11}, galactic outflows \citep{martin09, rubin10,coil11}, high--redshift \lya{} emitters \citep{Shibuya2014, Harish2021}, and high-redshift QSOs \citep{dawson04,hu04,Mazzucchelli2017}. Extensive work on the intergalactic medium \citep{casey08,lee18} and the circumgalactic medium of galaxies \citep{steidel10, werk12,rudie19, chen20} also uses LRIS.

LRIS has a multiobject capability enabled by machine-milled slit masks placed at the focal plane. Decades of observations have built a large archive of multislit observations available through the Keck Observatory Archive (KOA). These observations can in principle be used for science goals beyond the original programs, e.g. by applying new techniques to existing data or studying objects contained in science slits that were not of interest to the original observers. Unfortunately, the archive has no information on the sky positions of the slits in the mask, which were not stored in the LRIS observation headers (unlike other Keck spectrographs such as DEIMOS and MOSFIRE). This lack of information on slit locations means the archival spectra cannot be easily matched to objects on the sky, greatly diminishing the archival value of the data.

In this paper, we present \tilsotua, a code to calculate the sky positions of the science slits in LRIS slitmasks using archival mask design files used by Lick observatory's mask milling machines or the output files produced by \autoslit, the software used to design the slitmasks. We applied \tilsotua\ to the catalog of archived LRIS mask design files, available as of June 2020, to determine the position of every slit in every mask. With these slit positions, LRIS multislit observations stored in the KOA can be used by anyone with results from \tilsotua. Additionally, \tilsotua\ can be used by observers designing new slitmasks to store the slit geometries and display them directly on images. \tilsotua\ can also be used when reducing multislit LRIS data with {\tt PypeIt} \citep{Prochaska2020a,pypeit:zenodo} to assign sky coordinates to the extracted spectra. We distribute \tilsotua\ and the historical mask slit positions to the community via Github.\footnote{{\tt https://github.com/KeckObservatory/tilsotua}}

This description of \tilsotua\ is laid out as follows. In Section~\ref{sec:lrisfocalplane} we discuss the LRIS focal plane and slitmask details. In Section~\ref{sec:details} we lay out the details of the code. In Section~\ref{assessingperfomance} we characterize the quality of the slitmask reconstructions and final astrometry by comparison to Gaia coordinates. We summarize the results in section~\ref{sec:summary}.

\section{The LRIS Focal Plane and Mask System}
\label{sec:lrisfocalplane}

The LRIS $6\arcmin\times7.8\arcmin$ field of view can be used with longslit or multislit masks that sit at the telescope focal plane. The telescope focal surface is curved, and the slitmasks are located off of the optical axis. To account for this, the 0.41-mm thick aluminum masks are tilted in the dispersion direction, $X$, and bent in the cross-dispersion, $Y$, to create curved shape that more closely follows the telescope focal plane \citep[see Figure 5 of ][]{oke1995}. Mulitislit masks are designed by the user using the \autoslit\ code of \citet{cohen96}.\footnote{{\tt https://www2.keck.hawaii.edu/inst/lris/\linebreak autoslit\_{}WMKO.html}} The user mask designs are subsequently sent to the Lick Observatory, which operates the milling machine at Keck. The aluminum masks are cut by a mechanical milling machine \citep[see][for a description of the original machine]{oke1995}. The machines used to cut the masks have an accuracy better than 0.07 mm ($\approx0\farcs1$) in the $X$ direction and somewhat worse in the $Y$ direction \citep{oke1995}.

\begin{figure}
\centering
  \fig{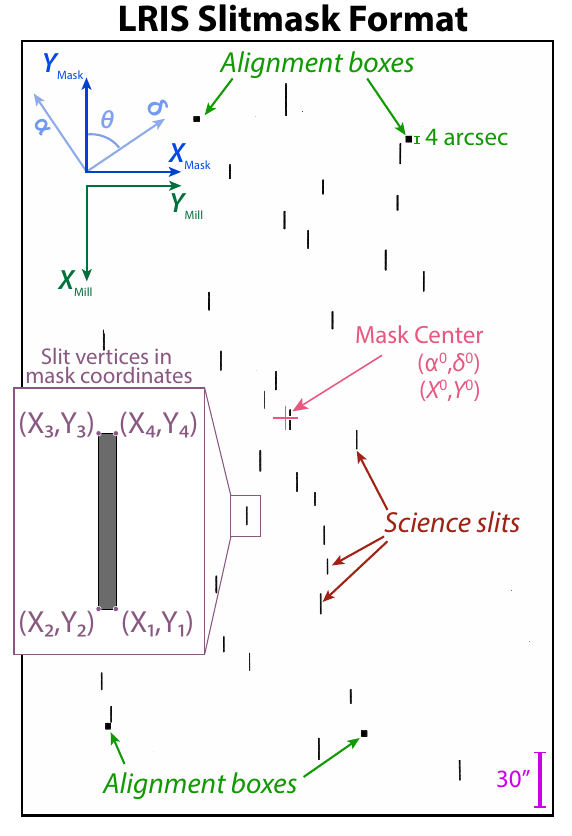}{1.0\columnwidth}{}

  \caption{An image of an illuminated slitmask (inverted grayscale) with pertinent features noted. We define several coordinate systems used to describe locations on the masks in the upper right (both physical and sky coordinates), and we give a scale bar along the bottom right of the image. We label several alignment boxes and science slits. The alignment boxes in this image are $4\arcsec$ on a side. The science slits are all $1\arcsec$ in width; an inset shows a close-up of one of the slits, labeling the vertices that are used to define the slit.  \label{fig:maskformat}}
\end{figure}

The user-designed masks must include both science slits and alignment boxes that allow the observer to align the mask precisely on the sky. Typical science slits are of order $1\arcsec$ in width and often greater than 10 arcseconds in length. The lengths are selected based on the need to place science slits on the mask without overlap in the cross-dispersion direction while providing good sky background measurements. The alignment boxes are typically $\ge4\,\arcsec$ on a side and are placed at the location of bright point-sources (with recommended magnitudes $15 < m < 19$) having well-known astrometry. Through-mask imaging is used to ensure the slitmasks are well-aligned on the sky, enabling the secure placement of science slits on very faint targets. We show in Figure~\ref{fig:maskformat} an annotated image of an illuminated multislit mask, labeling the two types of slits and the general geometry of the masks. 

Slit positions are defined in three different coordinate systems: sky coordinates, mask coordinates and milling machine coordinates (Figure~\ref{fig:maskformat}). The sky coordinates are the locations in celestial coordinates, RA and Dec, of the slit centers or slit vertices. Ultimately, the goal of \tilsotua\ is to produce the sky coordinate locations of the slits so that they can be compared with celestial objects. The rectangular mask coordinates, $(X_{\rm mask},Y_{\rm mask})$, denote the physical locations of the slits on the mask, defined such that the physical center of the mask is at $(X_{\rm mask}^{\rm 0},\, Y_{\rm mask}^{\rm 0}) = (305\ \rm mm,\, 0\ \rm mm)$. Given the physical size of the masks themselves, these coordinates extend over 174 mm $\leq X_{\rm mask}\leq$ 436 mm and -174 mm $\leq Y_{\rm mask} \leq$ 174 mm. The milling machine coordinate system, $(X_{\rm mill},Y_{\rm mill})$, is the coordinate system defined by the punch machine that physically perforates the aluminum masks. This rectangular system extends over 3.8 mm $\leq X_{\rm mill} \leq$ 351.8 mm and 1.3 mm $\leq Y_{\rm mill} \leq$ 263.3 mm with the mask center at $(X_{\rm mill}^{\rm 0},\, Y_{\rm mill}^{\rm 0}) = (177.8\ \rm mm,\, 132.3\ \rm mm)$. The mask and milling machine coordinate systems are rotated and flipped relative to one another (see axis definition in the upper left of Figure~\ref{fig:maskformat}); the transform between these coordinate systems is described in \S \ref{sec:details} below.

When designing masks, \autoslit\ is used to transform between sky coordinates and milling machine coordinates, as described in \cite{cohen96}. \autoslit\ intakes user-defined sky coordinates for a list of selected targets and defines slits to isolate the light from each celestial object. \autoslit\ then calculates and saves the milling machine coordinates of the slit vertices (its output {\tt .file3} file) for mask fabrication.  The user provides sky coordinates of the center of the mask and of the targeted celestial objects, and \autoslit\ uses these to determine slit locations and sizes. \autoslit\ calculates the slit positions in the milling machine frame to an accuracy of $\approx0\farcs05$. This transformation from sky to milling machine coordinates is a multistep process, which accounts for the complex projection of the sky onto the physical mask at the focal plane. 

In the first step of the coordinate transformation, \autoslit\ applies a correction to account for atmospheric refraction. Next, a gnomonic projection is applied to project the curved focal surface onto the mask. \tilsotua\ then rotates the projected coordinates on the mask surface to the mask coordinate system with the user selected position angle. Finally, a correction is applied for the telescope distortion at the focal plane as characterized by \cite{cohen07}. Once the distortion correction is applied, \autoslit\ finally calculates the positions of the slits in the milling machine frame to allow the cutting of the physical mask. We describe in Section~\ref{sec:details} the inversion of these steps applied by \tilsotua{}. Once the user finalizes the mask design, \autoslit\ writes three files covering the positions of the objects on the mask and the positions of the slits. The {\tt.file1} output contains the positions of the target objects supplied by the user. The {\tt.file2} output contains positions of the objects in the mask frame. The {\tt.file3} output contains the coordinates of the slit verticies in the milling machine frame. \autoslit\ also generates a DS9 region file of the mask design and an image file of the mask design. These files contain all the information needed to reconstruct the mask, but only the {\tt.file3} files were sent by users to Lick and archived.

The slit coordinate information output by \autoslit\ for the mask designs are not stored within the LRIS FITS headers or the Keck archive. This means that archive users do not have ready access to the slit positions, making using the multislit data for new science very difficult. However, the UCO/Lick Observatory created FITS file summaries of the designs submitted for mask milling (S. Allen 2020, private communication). We use this information to reconstruct the slit sky positions using \tilsotua{}, which will enable the use of the LRIS multislit data for new science goals. We give a brief summary of these archived mask design files, including information contained within each FITS extension, in Table~\ref{tab:fitsextensiondetails}. The archival FITS mask files contain the positions of the slit vertices in the milling machine frame (Figure~\ref{fig:maskformat}) along with the sky coordinates of the mask center, the equinox of the sky coordinates, and the sky position angle of each mask.

\section{tilsotua: Calculating Slit Sky Coordinates}
\label{sec:details}

 \tilsotua\ inverts each step performed by \autoslit\ and subsequently applies an astrometric correction to bring the mask positions into a modern astrometric frame, such as the Gaia frame \citep{GAIACollaboration2016b,GAIACollaboration2023j}. Until that astrometric correction, \tilsotua\ either directly inverts the equations used in each step of \autoslit's calculation or uses interpolation of complex corrections to invert the work of \autoslit{}.

\tilsotua\ accepts inputs of FITS files from the LRIS mask archive. Table~\ref{tab:fitsextensiondetails} contains details on the information found in each of the mask FITS file extensions. \tilsotua\ can also accept the output files created by \autoslit\ (namely the {\tt.file1} and {\tt.file3} output files) as inputs. Providing the {\tt.file1} and {\tt.file3} output files allows for additional information about the mask target objects to be added to the mask FITS files. This is the only way to fully populate every extension in the output FITS files. We refer to the input file to \tilsotua{}, regardless of format, as the mask design file throughout this paper. Each mask design file contains the milling machine coordinates for each individual slit vertex (Figure~\ref{fig:maskformat}). We refer to these collectively as the slit positions, with each slit corner treated as its own location on the mask, unless otherwise specified. 

\begin{figure}
\centering
    \includegraphics[scale=0.5]{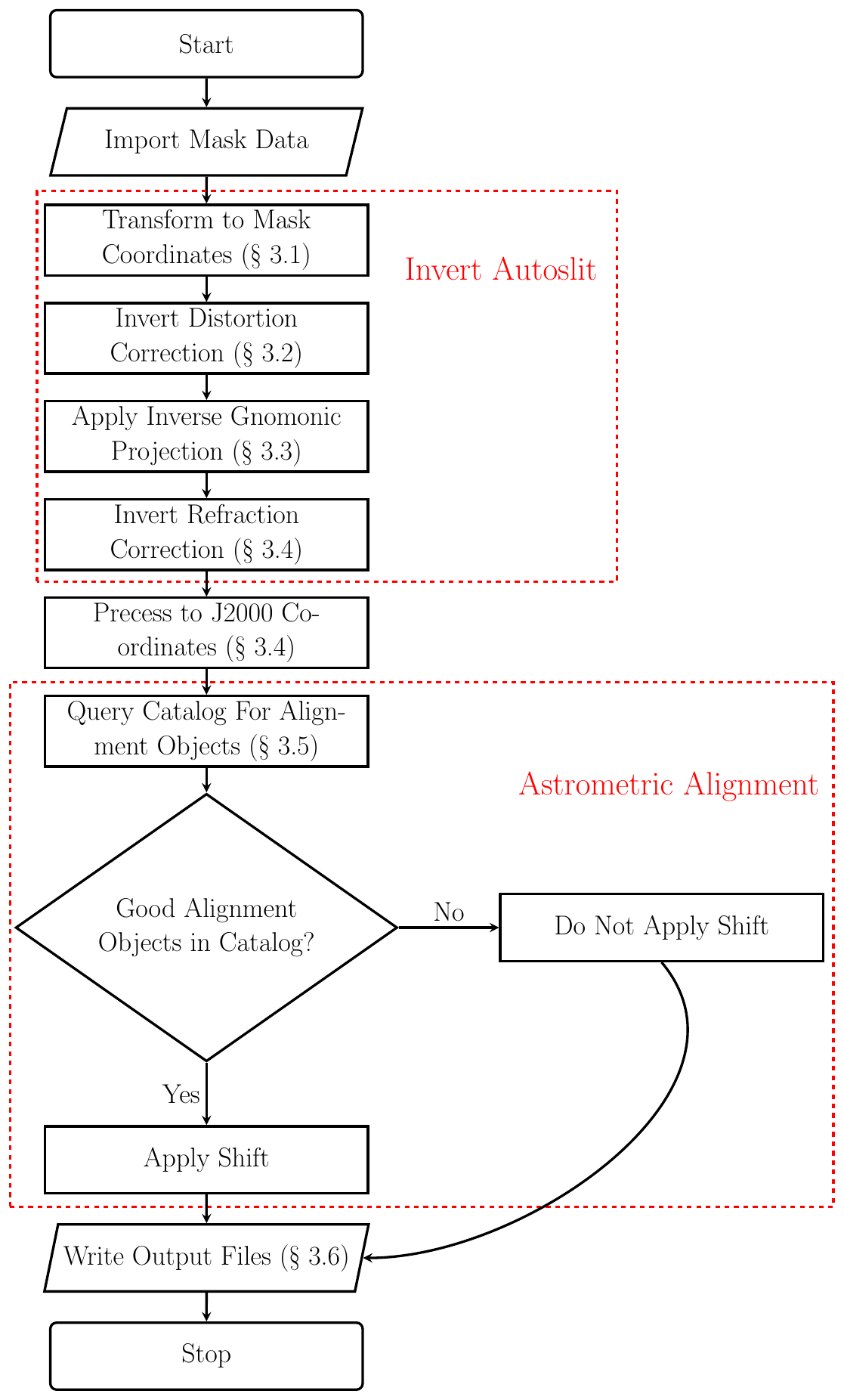}
    \caption{Flowchart of \tilsotua{}, showing each of the main steps. Through precession to J2000 coordinates, each step is applied to all masks. \tilsotua\ will not apply shifts for astrometric alignment, however, in cases where the shift does not significantly improve mask astrometry or catalog objects are not confidently identified for association with mask alignment boxes.\label{fig:flowchart}}
\end{figure}

\tilsotua\ follows the flow shown in Figure~\ref{fig:flowchart}. The code transforms the slit positions from milling machine to mask coordinates (\S~\ref{subsec:xycoords}), reverses the corrections for focal distortion and gnomonic projection applied by \autoslit\ (\S~\ref{subsec:xytowcs}), uses the platescale and known mask center sky coordinates to calculate initial sky coordinates for the slit positions, precesses the resulting positions to J2000 coordinates (\S~\ref{subsec:refractionandprecession}), and finally calculates and applies a shift of the sky positions to bring the mask astrometry into agreement with modern astrometry of the chosen reference catalog (\S~\ref{sec:shifts}). Table~\ref{tab:variables} summarizes variables used during the previous steps. The final coordinates of the slit centers, as well as the Gaia catalog targets associated with alignment boxes, are written to an output file and a FITS file with the Lick archive structure. \tilsotua\ creates a quick-look plot showing the calculated slit locations on the PanSTARRS \citep{flewelling20} image of the mask field, with cutouts showing close-ups of up to six alignment boxes. Finally, \tilsotua\ writes a DS9 region file of the mask slits. We describe each of these steps below.

\begin{deluxetable*}{llc}
\tablecaption{Reference Table of Relevant Coordinate Variables and Values\label{tab:variables}}
\tablehead{\colhead{Label} & \colhead{Description} &  \colhead{ Value(s)} }
\startdata
$X_{\rm Mask},Y_{\rm Mask}$ & Mask coordinates &  \nodata   \\
$X_{\rm Mill},Y_{\rm Mill}$ & Milling machine coordinates  &     \nodata  \\
$X^{\rm 0}_{\rm Mask},Y^{\rm 0}_{\rm Mask}$ & Mask center (Mask coordinates) & (305 mm, 0 mm) \\
$X^{\rm 0}_{\rm Mill},Y^{\rm 0}_{\rm Mill}$ & Mask center (Milling machine coordinates) & (177.8 mm, 132.3 mm) \\
$\theta$ & Sky position angle of mask & \nodata \\
$\phi$ & Angle of mask in focal plane in $X_{\rm Mask}$& \nodata \\
$\Phi$ & Bend angle of mask in $Y_{\rm Mask}$& \nodata\\
$X,Y$ & Tilt and bend corrected mask coordinates &  \nodata   \\
$N_{\rm pix}$ & Number of CCD pixels per side & 2048 \\
$p_s$ & Mask focal plane plate scale &  $0.7243\, {\rm mm}/\arcsec$ \tablenotemark{a}\\
$\alpha_0,\delta_0$ & Mask center sky coordinates & \nodata \\
$\alpha_0^{'},\delta_0^{'}$ & Mask center sky coordinates (refracted) $(\S~\ref{subsec:xytowcs})$ & \nodata \\
$X_{\rm CCD},Y_{\rm CCD}$ & Mask position in CCD pixels & \nodata \\
$X_{\rm Mask}^{\rm Undist},Y_{\rm Mask}^{\rm Undist}$ & Mask coordinates (pre-distortion correction) (\S~\ref{subsec:distortioncorrection}) & \nodata \\
$\Delta_{\rm dist,x},\Delta_{\rm dist,y}$ & Distortion corrections in x,y at $X_{\rm Mask},Y_{\rm Mask}$ (\S~\ref{subsec:distortioncorrection}) & \nodata \\
\enddata
\tablenotetext{a}{$p_{s}=0.7253 \ {\rm mm}/\arcsec$ prior to ADC use in Fall 2007}
\end{deluxetable*}

\subsection{Milling Machine to Mask Coordinate System Transformation}
\label{subsec:xycoords}

The mask design files contain the positions of the vertices for each slit in milling machine coordinates, ($X_{\rm Mill},Y_{\rm Mill})$. However, \autoslit\ works in the mask coordinate system, ($X_{\rm Mask},Y_{\rm Mask})$, and only uses the milling machine coordinates in the final {\tt .file3} output file. After ingesting the input file, \tilsotua\ first converts from the milling machine coordinates frame to the mask coordinates using the conversions
\begin{eqnarray}\label{eq:xcoordconversion}
X_{\rm Mask} = Y_{\rm Mill}+ Y^{\rm 0}_{\rm Mill}, \nonumber \\
Y_{\rm Mask} = -X_{\rm Mill}+X^{\rm 0}_{\rm Mill}.
\end{eqnarray}
\noindent Here $(X_{\rm Mill}^{\rm 0},Y_{\rm Mill}^{\rm 0})$ is the center of the mask in the milling machine frame (given in mm) as given in Table~\ref{tab:variables}.

\subsection{Reversing the Distortion Correction}
\label{subsec:distortioncorrection}

After transforming to mask coordinates, \tilsotua\ inverts the LRIS focal plane distortion correction made to the $(X_{\rm Mask},Y_{\rm Mask})$ positions by \autoslit. This correction accounts for telescope distortions at the (curved) LRIS focal plane (\citealp{cohen96}; see also \citealp{cohen07} for updates). The original \autoslit\ calculation is contained in its \slitastrometry\ subroutine, the details of which are described in our Appendix~\ref{sec:Details of slit astrometry Routine}. 

The equations used in \autoslit{}'s distortion correction calculation, Equations~\ref{eq:XCCD} through \ref{eq:YCCD}, cannot be readily inverted. To reverse the distortion correction, \tilsotua\ uses \autoslit{}'s \slitastrometry\ routine to calculate the correction applied during this step at each point in the mask and then reverses that (approximate) correction. To do so, \tilsotua\ generates a $300\times 300$ grid of positions over the nominal mask field, which corresponds to the uncorrected positions in the mask coordinate system, $(X_{\rm Mask}^{\rm Undist},Y_{\rm Mask}^{\rm Undist})$. \tilsotua\ calculates associated distortion correction for each grid location, and uses these corrections to create a new grid of distortion-corrected positions in $(X_{\rm Mask},Y_{\rm Mask})$. Then, \tilsotua\ applies a linear interpolation scheme over the new grid to obtain a continuous correction, $(\Delta_{\rm dist,x},\Delta_{\rm dist,y})$, over position on the mask. By generating this interpolation as a function of the distortion-corrected positions, \tilsotua\ can determine the correction applied at each of the slit positions. \tilsotua\ subtracts the applied distortion correction value from each slit position $(X_{\rm Mask},Y_{\rm Mask})$ to obtain $(X_{\rm Mask}^{\rm Undist},Y_{\rm Mask}^{\rm Undist})$ for the slits:

\begin{eqnarray}
\label{xdistcorr}
  X_{\rm Mask}^{\rm Undist} = X_{\rm Mask}-\Delta_{\rm dist,x}, \nonumber \\
\label{ydistcorr}
  Y_{\rm Mask}^{\rm Undist} = Y_{\rm Mask}-\Delta_{\rm dist,y}.
\end{eqnarray}

\noindent These coordinates represent the mask coordinates prior to the application of the distortion correction by \autoslit{}.

\subsection{Transforming Physical Mask Coordinates to Sky Coordinates}
\label{subsec:xytowcs}

After removing the applied optical distortion corrections, \tilsotua\ converts the physical mask coordinates to sky coordinates (RA and Dec, denoted as $\alpha$ and $\delta$) for each slit position. To complete this calculation, \tilsotua\ uses $(X_{\rm Mask}^{\rm Undist},Y_{\rm Mask}^{\rm Undist})$ along with the central sky coordinate of the mask, the sky position angle (E of N), the Keck focal plane plate scale, and an inverse gnomonic projection. First, \tilsotua\ calculates $(X,Y)$, correcting the $(X_{\rm Mask}^{\rm Undist},Y_{\rm Mask}^{\rm Undist})$ positions corrected for the angle of tilt, $\phi$, about the x-axis of the mask and angle of bend, $\gamma$, about the y-axis \citep{cohen96} \cite[see also][Figure 5]{oke1995}:
\begin{eqnarray}
\label{maskangle}
X = \cos (\phi) X_{\rm Mask}^{\rm Uncorr} \nonumber, \\
\label{bend}
Y= \cos (\gamma) Y_{\rm Mask}^{\rm Uncorr}.
\end{eqnarray}
\noindent \tilsotua\ then transforms mask-frame $(X,Y)$ coordinates into angular positions relative to the Mask coordinate system origin, $(\eta,\nu)$, aligned with the sky at the focal plane, accounting for mask position angle, $\theta$, and the telescope plate scale, $p_s$:

\begin{eqnarray}
\label{eq:eta}
\eta =  \frac{\cos(-\theta)X-\sin(-\theta)Y}{206265 \times p_{s}}, \nonumber \\
\label{eq:nu}
\nu =  \frac{\sin(-\theta)X+\cos(-\theta)Y}{206265 \times p_{s}}.
\end{eqnarray}

\noindent The plate scale varies depending on whether the mask was designed to be used with the atmospheric dispersion corrector (ADC). Numerical values for $p_{s}$ are contained in Table~\ref{tab:variables}. Note that the angular coordinates, $\eta$ and $\nu$, are in radians.

Now \tilsotua\ converts from coordinates in the mask focal plane to sky coordinates. \autoslit\ performs a gnomonic projection to map the curved surface of sky coordinates to positions on the flat mask. \tilsotua\ inverts this step using the inverse gnomonic projection:

\begin{eqnarray}
\label{ra}
{\rm \alpha^{'}} =  & \, {\rm \alpha}_{\rm tan}^{'}+  \tan^{-1}\left(\frac{\eta \sin{c}}{\rho \cos{\rm \delta}_{\rm tan}^{'}\cos{c}-\nu \sin{{\rm \delta}_{\rm tan}^{'}}\sin{c}}\right)\nonumber, \\
\label{dec}
{\rm \delta^{'}} = & \,  \sin^{-1}\left(\cos{c} \sin{{\rm \delta}_{\rm tan}^{'}}+      \frac{\nu \sin{c} \cos{{\rm \delta}_{\rm tan}^{'}}}{\rho}\right),
\end{eqnarray}

\noindent where $(\alpha^{'},\delta^{'})$ are slit positions in sky coordinates at the telescope aperture. In these equations $\rho$ is the distance from the mask center, 
\begin{equation}
\label{rho}
\rho = (\eta^2+\nu^2)^{\frac{1}{2}},
\end{equation}
\noindent and $c$ is the angular distance from the mask center,
\begin{equation}
\label{c}
c = \tan^{-1}\rho . 
\end{equation}

\noindent The position $(\alpha_{\rm tan},\delta_{\rm tan})$ is the tangent point of the gnomonic project. This point is calculated by shifting the mask center sky coordinates, $(\alpha_0^{'},\delta_0^{'})$ to the position of the origin of the mask coordinate system:

\begin{eqnarray}
\label{ra0correction}
\alpha_{\rm tan} = \alpha_0^{'}-  \frac{X^{\rm 0}_{\rm Mask}}{p_s} \frac{\cos( -\theta) }{\cos \delta_0^{'}}, \nonumber \\
\label{dec0correction}
\delta_{\rm tan} = \delta_0^{'} - \frac{X^{\rm 0}_{\rm Mask}}{p_s}\sin(-\theta).
\end{eqnarray}
\noindent Here $X_{\rm Mask}^{\rm 0}$ is the x coordinate of the mask center in the mask coordinate system (Table~\ref{tab:variables}). The coordinates $(\alpha_0^{'},\delta_0^{'})$ are sky coordinates after an atmospheric refraction correction is applied (see \S~\ref{subsec:refractionandprecession}), not the recorded central sky coordinate input by the user into \autoslit\ when the mask was initially designed. The tangent point position, $(\alpha_{\rm tan},\delta_{\rm tan})$, is used when inverting the gnomonic projection.

The coordinates $(\alpha^{'},\delta^{'})$ are not yet the celestial coordinates of the slits, as the effects of atmospheric refraction have not been removed.

\subsection{Reversing the Refraction Correction and Precessing Coordinates}
\label{subsec:refractionandprecession}

To recover the slit sky coordinates, \tilsotua\ removes the correction for atmospheric refraction applied to transform from sky coordinates to $(\alpha^{'},\delta^{'})$. \autoslit\ calculates the refraction effect using the recorded hour angle typical of observation, a central wavelength, the latitude of the Keck Observatory, and standard atmospheric temperature and pressure. \tilsotua\ uses \autoslit{}'s refraction correction routine to calculate a correction grid over the extent of the mask, assuming the default values for wavelength (6000 \AA), temperature ($0^\circ$ C), and pressure (486 mm of Hg). \tilsotua\ interpolates between the grid points to calculate the effect at each slit vertex, in the same manner as in \S~\ref{subsec:distortioncorrection}, using a 2D linear interpolation scheme. \tilsotua\ subtracts the shifts applied to each $(\alpha^{'},\delta^{'})$ slit position to recover the pre-refraction, true sky positions, $(\alpha,\delta)$.

The $(\alpha,\delta)$ positions calculated by \tilsotua\ are equivalent to the original slit sky positions used when the mask was designed. As a final step, \tilsotua\ precesses the slit positions to the J2000 equinox from the original user's chosen frame using the {\tt astropy} precession function, which we found to be almost identical to that used by \autoslit{}.

\subsection{Aligning with Modern Astrometry}
\label{sec:shifts}

While our code reproduces the slit locations created by the original \autoslit\ users (demonstrated through ``full circle'' algorithm testing described in Section~\ref{fullcircle}), this will not necessarily ensure the slit positions align with objects on the sky. The quality of the alignment depends on the quality of the absolute astrometry used in the original mask design. Slit positions from our calculations sometimes show systematic offsets from objects on the sky in surveys using modern astrometry; notably, we find the square alignment boxes can be consistently offset from the bright, easily-identifiable objects on which they should be centered. A majority of masks with clear offsets show alignment boxes systematically offset from the targets in a way that is consistent with a simple linear shift relative to modern astrometry. We have built functionality into \tilsotua\ to identify likely alignment objects in modern catalogs (PanSTARRS, Gaia, or user provided) and apply linear shifts to all slit positions to best center the alignment objects within the boxes.

To generate the list of targets associated with the alignment boxes, \tilsotua\ queries a modern reference catalog and obtains the positions of objects near each alignment box. By default this catalog is Gaia DR3, but users can choose instead to use the PanSTARRS DR2 or custom user catalogs. The catalog object closest to the center of each alignment box is chosen as the object intended to be associated with that box. \tilsotua\ does not make magnitude cuts when selecting catalog objects during this step, as mask creators at times use objects fainter than those recommended by the observatory (e.g., when observing high-latitude fields with few bright Galactic stars). 

\tilsotua\ calculates the shift between each alignment box center and the catalog position of its target, and subtracts the average of these shifts derived from the alignment boxes to all slit positions. During this step, \tilsotua\ checks for alignment object mismatches (which can happen, e.g., when observers used faint galaxies not detected by the catalog survey). Since the shifts are assumed to be linear, \tilsotua\ adopts an outlier rejection to remove misidentified alignment box objects from the shift calculation.

To determine outliers, \tilsotua\ identifies the  individual alignment box shifts likely due to incorrectly identified objects with a general random sample consensus (RANSAC) method for outlier identification. \tilsotua\ uses the PyRANSAC package\footnote{https://github.com/adamlm/pyransac} to apply the RANSAC algorithm to a linear model of the set of individual box shifts in both RA and Dec. The RANSAC algorithm identifies outliers and inliers from the provided shifts by determining an optimal fit of a flat linear model of the box offsets. The first outliers are classified from fitting the RA offsets. After excluding the already-identified outliers, the Dec offsets are fit and any additional outliers are discarded. \tilsotua\ calculates the shift value in RA ($\Delta_{\alpha}$) and Dec ($\Delta_{\delta}$) as the average of the remaining individual offsets. For our testing described in Section~\ref{sec:skyperformance}, we also record $\Delta_{\theta}$, the total angular offset. The box-target offsets, ($\Delta_{\alpha}$,$\Delta_{\delta}$), are then removed from all sky slit positions in the mask. This method generally does a good job identifying which box offsets are outliers due to misidentified objects, however the extremely small samples sizes for alignments targets in an individual mask are limiting. Due to this, the RANSAC method can if rerun return slightly different results for which shifts belong to the outlier group. Variation in outlier selection can slightly affect the final shift applied to the mask. We suggest users carefully check the final results once outliers are identified using the quick-look plots to confirm the final astrometric agreement is satisfactory.

In many cases, by removing outliers, \tilsotua\ calculates a shift that brings the mask into good agreement with modern astrometry. However, we also encounter masks where applying the calculated shift does not improve the overall agreement between the alignment box centers and the chosen catalog object positions. Such masks have large dispersions in the individual box-target offsets, even after applying outlier rejection. \tilsotua\ does not apply a shift to masks in cases where the dispersion in the calculated offsets is $>1\arcsec$ or in cases where the ratio of the average offset to the dispersion is $<0.7$. For these cases where no shift is applied, we recommend users check the mask alignment and consider creating a custom catalog of alignment targets, e.g., derived from pre-imaging data in the KOA.

We investigated the minimum  number of alignment targets needed to obtain a robust shift that improves the absolute astrometric accuracy of the masks. Based on experiments with numerous masks, we find that three objects will give a reliable shift for a mask, and even two objects will give a usable shift in most cases. This is understandable given our intrinsic model is a simple linear shift, but more objects hedge against mis-identifications that may bias the calculated offset. The applied shifts are discussed further in \S~\ref{sec:skyperformance}.

\subsection{Description of \tilsotua\ Output}
\label{subsec:codeoutputs}

The \tilsotua\ results are saved in four files: 1) a FITS file in the same structure as the archival mask FITS files (matching the structure for DEIMOS \citep{Faber2003}) with recorded $(\alpha,\delta)$ slit positions; 2) a comma-separated value (CSV) file that contains the slit sky and mask positions and the associated alignment target information; 3) a {\tt SAOImage DS9}\footnote{{\tt https://sites.google.com/cfa.harvard.edu/\linebreak saoimageds9/home}} region file for displaying the slits over images; and 4) a ``quick-look'' plot in PDF format showing the on-sky positions of the slits compared with PanSTARRS imaging \citep{flewelling20}.

In the mask FITS files from the UCO/Lick archive, the {\tt DesiSlits} extension contains slit heights, widths, position angles, and empty columns to store the central sky positions of the slits. \tilsotua\ populates that information and stores the final $(\alpha$,$\delta)$ positions for each of the slit centers. Additionally, to provide more information to interested users, \tilsotua\ also saves the $(X_{\rm mask},Y_{\rm mask})$ and $(\alpha, \delta)$ positions of the slit vertices and slit centers in the CSV-file.

\begin{figure*}
\centering
  \includegraphics[scale = 1.2]{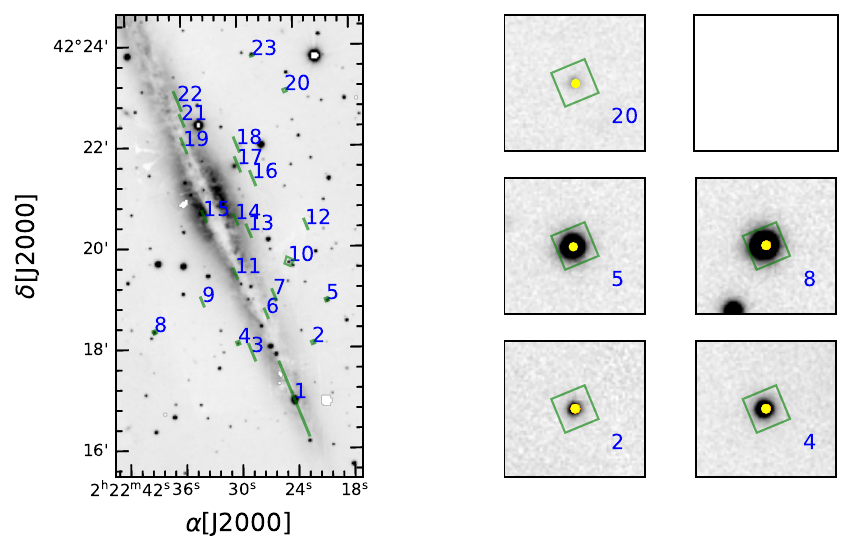}
  \caption{Sample ``quick-look'' plot showing the reconstructed LRIS slit mask on a PanSTARRS \citep{flewelling20} image. This mask targeted objects centered on the nearby edge-on galaxy NGC 891 (mask PI: Howk). The left image shows a PanSTARRS $g$-band image of the full mask field with all slits displayed in green and marked with their corresponding number in the mask FITS file (ordered according to their $Y_{\rm Mask}^{\rm Uncorr}$ coordinates). In this case, the science slits have a width of $1\farcs0$ and a length $\ge 10\arcsec$, while the alignment boxes are $4\,\arcsec\times4\,\arcsec$. Each of the smaller image cutouts on the right shows a $10 \arcsec$ region centered on up to six alignment boxes. The positions of the catalog sources identified as the alignment targets are marked in yellow. The magnitude of the first order astrometric correction (shift) applied to bring this mask onto the Gaia frame from the user astrometry was $|\Delta_{\rm \theta}| = 0\farcs35\pm0\farcs12$ (std. dev.).   \label{ngc891results}}
\end{figure*}

We have calculated the positions of slits for {\em all} of the archival mask design files available through UCO/Lick as of June 2020. As all of these are outside of the KOA proprietary period, we distribute the \tilsotua\ output for these masks with the code. An example of our mask solutions is shown with the output quick-look plot in Figure~\ref{ngc891results}. The slit positions for a mask targeting the edge-on galaxy NGC 891 (mask PI: J.C. Howk) are displayed on a PanSTARRS $g$-band image of the field. The left image shows all mask slits, numbered from bottom to top of the mask. This numbering system is useful for connecting individual slits to objects in science exposures. The inset images on the right of the plot show enlarged views of $10\,\arcsec\times10\,\arcsec$ regions around up to six alignment boxes. These panels show the alignment boxes, with the targets identified by our shift algorithm marked with yellow dots. While the science slits often target faint objects that may not be detected in PanSTARRS imaging, the alignment boxes usually contain bright objects (recommended magnitudes $15 < m < 19$); the cutouts on the right of this figure offer the most straightforward way to assess if the mask solution is well-aligned with objects on the sky.

\section{Assessing Slit Reconstruction Performance}
\label{assessingperfomance}

There are two potential sources\footnote{Proper motion of objects could also be a source of error, especially for older masks, but we do not consider the impact of proper motions here.} of astrometric discrepancies between our code output and the positions of science and alignment targets on the sky: 1) errors in the inversion of \autoslit\ calculations, and 2) errors in the user-provided astrometry used to create the masks. For example, in the field shown in Figure \ref{ngc891results}, the final alignment boxes are well centered on the alignment targets. In this case, the central position of the mask after inverting the \autoslit\ calculations was shifted by $|\Delta_{\rm \theta}|=0\farcs35\pm0\farcs12$ to bring the positions of the alignment boxes calculated by our code (the reconstruction of the input astrometry) into agreement with the positions of the alignment targets identified in the Gaia astrometric catalog. The RMS offset of the final alignment box centers from their targets for the mask in Figure~\ref{ngc891results} is $0\farcs11$, which is a combination of the potential sources of error. This is typical of masks designed with relatively modern input astrometry. In this section we characterize the typical errors associated with inverting \autoslit\ (\S~\ref{fullcircle}) and with the full reconstruction, including user-input astrometry errors (\S\ref{sec:skyperformance})

\subsection{Full Circle Testing Against \autoslit }
\label{fullcircle}

To assess the contribution to the uncertainties in the final slit positions from errors in our reversal of the \autoslit\ algorithm, we performed a ``full-circle'' test. This test involves calculating sky slit positions of archival masks using \tilsotua{}, without adjusting to modern astrometry. Then, we recalculate the $(X_{\rm mask},Y_{\rm mask})$ positions using 
\autoslit{}. We then compare the mask archival input positions to the recalculated mask positions to check the level of agreement between them. If \tilsotua\ perfectly inverted \autoslit\, these positions would be identical. We selected 250 random archived masks and calculated sky positions of the slits in those masks. We ran \autoslit\ on input files to obtain the recalculated slit vertex positions in the mask milling machine coordinate system. We find that the mask coordinates of the original and reconstructed slit vertex positions agree to within a thousandth of a mm, with a total RMS offset of $9 \times 10^{-4}\,\rm{nm}$ (i.e., $0\farcs001$) over the 27280 individual slit vertices used in our tests. This level of agreement between the original mask input astrometry and our output astrometry after inversion of the \autoslit\ calculations implies that the reconstruction process contributes negligibly to the total error.

\subsection{Characterizing the On-Sky Performance}
\label{sec:skyperformance}

Our ``full-circle" testing demonstrates that \tilsotua\ accurately reproduces the user-input astrometry. However, the quality of the input astrometry adopted by the original users varies. There are often clear, systematic offsets between the initial \tilsotua -produced locations of the slit positions and the locations of catalog objects on the sky (e.g., from Gaia or PanSTARRS). We chose to correct the input astrometry using a simple first-order correction, shifting the masks so that the alignment boxes are better centered on their likely alignment targets (\S \ref{sec:shifts}).

\begin{figure*}
\centering
  \includegraphics[scale = 1.1]{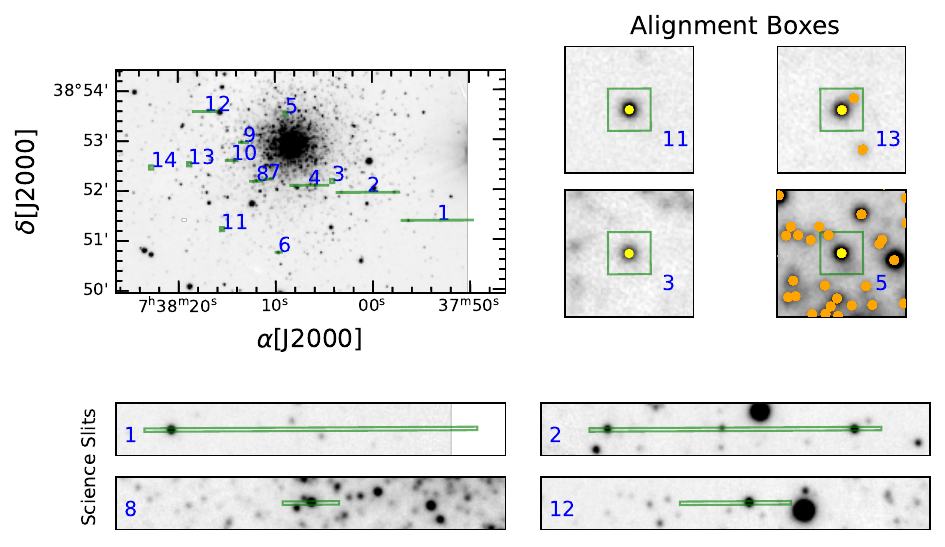}
  \caption{Modified ``quick-look'' plot showing the reconstructed LRIS slit mask on a PanSTARRS \citep{flewelling20} image. This mask targeted objects centered on the globular cluster NGC 2419 (mask PI: J. Cohen). Plot details match Figure~\ref{ngc891results}. Four Additional $100\arcsec \times 14\arcsec$ cutout images hightlight science slits. Additional Gaia catalogs are marked in orange. The magnitude of the first order astrometric correction (shift) applied to bring this mask onto the Gaia frame from the user astrometry was $|\Delta_{\rm \theta}| = 0\farcs15\pm0\farcs04$ (std. dev.).   \label{ngc2419results}}
\end{figure*}

Figures~\ref{ngc891results} and \ref{ngc2419results} show the results of applying \tilsotua\ to two archival slitmasks. Figure~\ref{ngc891results} shows the \tilsotua\ ``quick-look'' plot for a mask targeting objects around the galaxy NGC 891, while Figure~\ref{ngc2419results} highlights a mask placed on the globular cluster NGC 2419. In both cases, the level of agreement between the final mask astrometry and the Gaia object positions is excellent. Figure~\ref{ngc2419results} in particular emphasizes the ability of \tilsotua{}'s outlier rejection to remove incorrect targets associated with an alignment box even in denser fields. The NGC2419 mask alignment box cutout images in Figure~\ref{ngc2419results} show the final associated objects in yellow. The final alignment of the mask, after a shift of $\Delta_{\rm \theta}=0\farcs154$, aligns science slits with stars which can be seen in the PanSTARRS image and centers objects in the alignment boxes. This confirms that these objects were correctly identified and the many nearby objects (shown in orange in the cutouts, notably in alignment boxes 5 and 13) were not mistakenly selected. In addition to the alignment boxes shown in the standard quick-look plots, Figure~\ref{ngc2419results} shows cutouts of four of the science slits targeting stars in the globular cluster itself. These slits have widths of 1\arcsec, and the figure demonstrates that the science targets are reliably placed within these slits by the \tilsotua\ calculations.

\begin{figure}
\centering
  \includegraphics[scale=0.5]{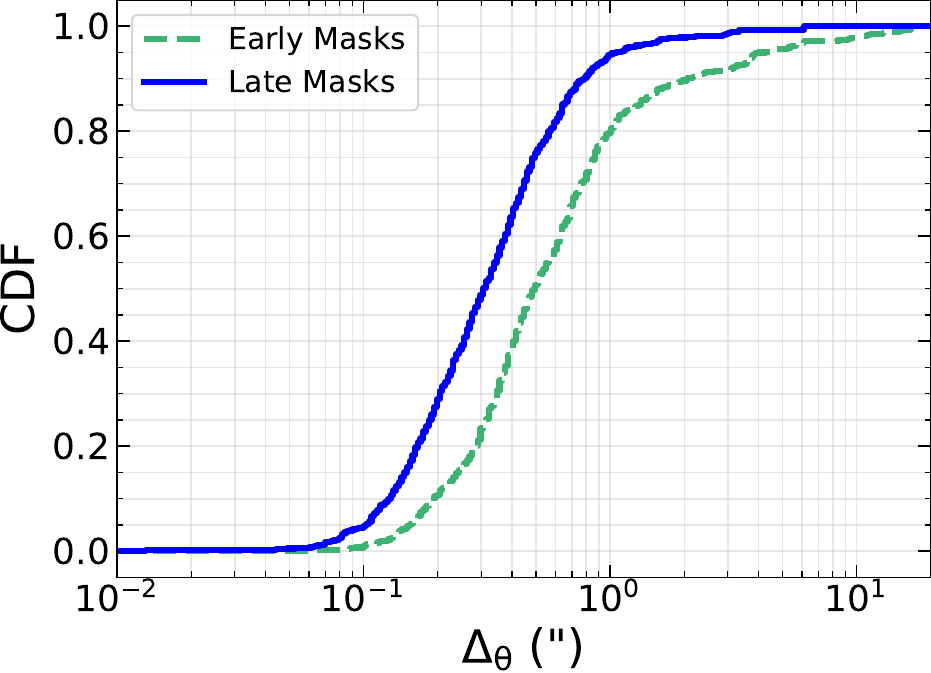}
  \caption{The cumulative distribution function of the magnitude of the astrometric shifts applied to a sample of 3,207 masks to bring the final astrometry into agreement with the Gaia astrometric frame. These corrections were calculated by minimizing the offsets between the centers of the mask alignment boxes and the positions of presumed alignment targets in the Gaia EDR3 catalog. Newer masks (blue) on average require significantly smaller astrometric corrections than older masks (green). \label{fig:maskshiftscdf}}
\end{figure}

For the 3,207 science masks in the UCO/Lick archive, we tracked the positions of the alignment box centers compared to the identified target objects before and after applying the astrometric correction to the mask. We show in Figure \ref{fig:maskshiftscdf} the cumulative distribution of linear shifts, $\Delta_{\rm \theta}$, applied to minimize the offsets between the alignment boxes and the predicted alignment targets. We include only the 62\% of masks for which shifts were applied, i.e., the masks for which the dispersion in the calculated shifts for all alignment boxes was small compared to the mean. We separately show the results for masks designed before 2010 (early) and from 2010 through today (late). There is a slight improvement in the input astrometry with time, but the majority of masks require small shifts over both periods. We find $>94\%$ and $>78\%$ of masks required corrections smaller than $1\arcsec$ for the late and early times, respectively. Of the remaining masks for which the corrections were not applied due to the large dispersion in their individual alignment box shifts, the mean shift implied (but not applied) was $\leq 1\arcsec$ in $66\%$ of the cases. These cases likely still include outliers due to misidentified alignment targets.

Table~\ref{tab:shiftstats} summarizes the statistics of the shifts applied to the masks in the early and late time periods. Using only correctly identified objects, the early masks have a total median offset between the alignment box centers and identified Gaia objects of $0.350\,\arcsec$ before bringing the mask results into agreement with the Gaia astrometry. After applying the shifts to each mask, the total median offset between alignment box centers and Gaia objects is $0\farcs10$. The late mask sample astrometry is similar to the early masks (Figure~\ref{fig:maskshiftscdf}), with a total median offset of $0\farcs33$ before and $0\farcs09$ after bringing the final astrometry into alignment with Gaia. The magnitude of the shift applied to bring a mask into agreement with Gaia astrometry does change slightly between the early and late masks, decreasing from $0\farcs51$ for the early masks to $0\farcs31$ for the late masks.

\begin{figure}
\centering
  \includegraphics[scale=0.45]{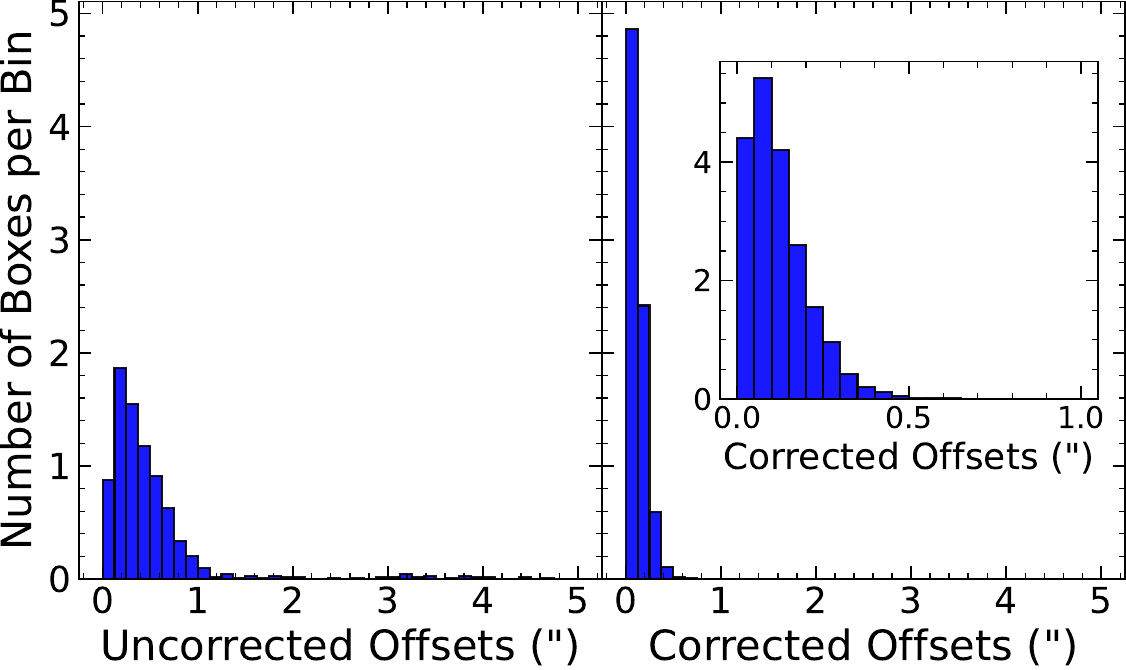}
  \caption{Comparison of the offsets between the alignment box centers and the identified alignment targets from the Gaia catalog before (left) and after (right) applying the astrometric correction calculated by \tilsotua. The inset on the right shows a finer view of the distribution within $\leq 1\arcsec$ (note the smaller bin sizes than the main figure).  
  \label{fig:preandpostshiftoffsets}}
\end{figure}

Figure~\ref{fig:preandpostshiftoffsets} shows the distribution of angular offsets between the alignment box centers and their associated targets from the Gaia DR3 catalog before and after the astrometric correction is applied to bring the mask results into agreement with Gaia astrometry. There is clear overall improvement after applying the astrometric shifts in the agreement between the alignment boxes and their presumptive targets. The corrected positions of the alignment boxes are largely within a few tenths of arcseconds of the presumed targets. This level of absolute astrometric alignment is sufficient to identify the objects included in the science slits, typically having $\sim1\arcsec$ width. 

\begin{figure}
\centering
  \includegraphics[scale=0.75]{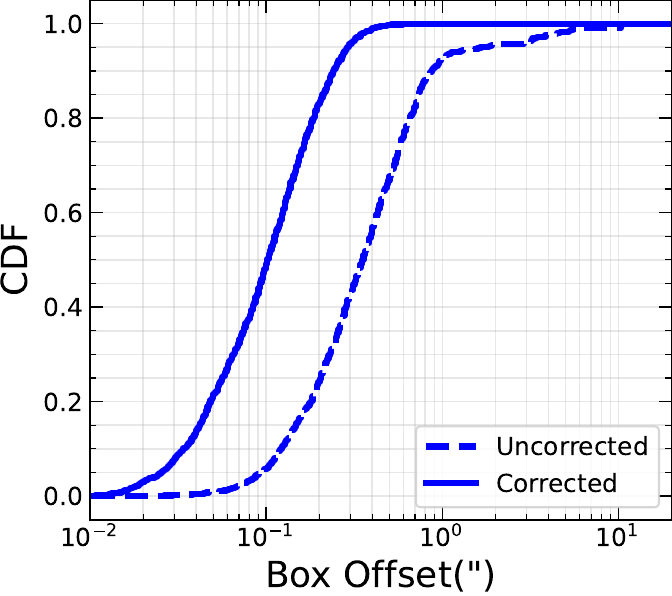}
  \caption{The cumulative distribution functions of the total offset values between the mask alignment box centers and associated Gaia objects as shown in Figure~\ref{fig:preandpostshiftoffsets} before and after the slit positions are corrected. The box offset values greatly improve once the calculated shift based on catalog object positions is applied.\label{fig:totaloffsetcdf}}
\end{figure}

To highlight the improvement in mask astrometric alignment, Figure~\ref{fig:totaloffsetcdf} directly compares the cumulative distribution of offsets between the alignment box centers and Gaia objects before and after the final shift is applied. This figure also shows how drastic the improvement in agreement with modern astrometry is with the shift applied, with final alignments generally within tenths of an arcsecond. Across all masks, $\sim80\%$ of the offsets remaining after the astrometric corrections are $<0\farcs2$. After that improvement, we are satisfied with the final level of agreement between mask and Gaia astronomy.

\begin{deluxetable*}{lccc}
\tablecaption{Summary Statistics of Astrometric Offsets\label{tab:shiftstats}}
\tablehead{\colhead{Timeframe} & \colhead{$\Delta$ ($\arcsec$)\tablenotemark{a}} &  \colhead{Pre-Correction} & \colhead{Post-Correction} \\ \colhead{} & \colhead{} &  \colhead{Box-Object Offsets ($\arcsec$)} & \colhead{Box-Object Offsets ($\arcsec$)}}
\startdata
$< 2010$ & $0.51^{+0.91}_{-0.25}$ &  $0.35 ^{+0.37}_{-0.20}$ &  $0.10 ^{+0.10}_{-0.05}$ \\
$\geq 2010$ & $0.31^{+0.33}_{-0.16}$ & $0.33^{+0.39}_{-0.17}$  &  $0.09^{+0.10}_{-0.05}$ \\
\enddata
\tablenotetext{a}{Average and standard deviations of calculated shifts for the masks within the respective time frames}
\end{deluxetable*}

\section{Summary}
\label{sec:summary}

LRIS has a rich archive of multislit data that has been used for a wide range of science. This data is hard to use beyond the scope of the original projects without the sky positions of slits in the observation slitmask. While the sky positions were not stored alongside the observations, slitmask design information was archived at UCO/Lick. We present and make available on Github \tilsotua, which recalculates the sky positions of the mask slits. 

\tilsotua\ inverts the work of the \autoslit\ code used to create the mask designs. Using the archived UCO/Lick mask FITS files or the \autoslit\ output files from mask generation, \tilsotua\ calculates the positions of the mask slits in sky coordinates. \tilsotua\ transforms between the coordinate systems used for the mask milling machine and slitmask to sky coordinates while correcting for distortion at the focal plane and atmospheric refraction. A linear shift is then applied to bring the sky positions into agreement with modern astrometry as the final step. The results are written to output files which include the information necessary to reconstruct the slit mask on the sky and match archival spectra with the targeted objects.

We tested \tilsotua{}'s ability to reconstruct slit sky positions. We checked \tilsotua{}'s accuracy inverting \autoslit{}'s calculations with a ``full-circle" test. We ran a sample of masks through \tilsotua\ without correcting to modern astrometry, then pass the resulting slit sky position back through \autoslit{}. When we compare the original mask coordinates to the reconstructed positions, the average difference is on the order of $0\farcs001$.  When comparing the final positions of archival alignment boxes to modern Gaia astrometry, we find our results to be accurate to within $0\farcs2$ relative to the Gaia astrometry for $\sim80\%$ of slits. The median offset between the bright objects used to position the mask during observations and the alignment boxes created for them in the mask is $0\farcs09$ for late masks ($>2010$). We distribute the final positions for all archival mask slits (as of June 2020) from UCO/Lick as part of the \tilsotua\ Github repository.

\begin{acknowledgements}
Support for this research was made by NASA through the Astrophysics Data Analysis Program (ADAP) grant 80NSSC21K0648. We also acknowledge support for this program through NSF grant award number AST-1910255. Sunil Simha is currently supported by the joint Northwestern University and University of Chicago Brinson Fellowship.

We appreciate support and advice from Steve Allen of UCO/Lick and Luca Rizzi of the Keck Observatory.

The Pan-STARRS1 Surveys (PS1) and the PS1 public science archive have been made possible through contributions by the Institute for Astronomy, the University of Hawaii, the Pan-STARRS Project Office, the Max-Planck Society and its participating institutes, the Max Planck Institute for Astronomy, Heidelberg and the Max Planck Institute for Extraterrestrial Physics, Garching, The Johns Hopkins University, Durham University, the University of Edinburgh, the Queen's University Belfast, the Harvard-Smithsonian Center for Astrophysics, the Las Cumbres Observatory Global Telescope Network Incorporated, the National Central University of Taiwan, the Space Telescope Science Institute, the National Aeronautics and Space Administration under Grant No. NNX08AR22G issued through the Planetary Science Division of the NASA Science Mission Directorate, the National Science Foundation Grant No. AST-1238877, the University of Maryland, Eotvos Lorand University (ELTE), the Los Alamos National Laboratory, and the Gordon and Betty Moore Foundation.

This work has made use of data from the European Space Agency (ESA) mission {\it Gaia} (\url{https://www.cosmos.esa.int/gaia}), processed by the {\it Gaia} Data Processing and Analysis Consortium (DPAC, \url{https://www.cosmos.esa.int/web/gaia/dpac/consortium}). Funding for the DPAC has been provided by national institutions, in particular the institutions participating in the {\it Gaia} Multilateral Agreement.

\end{acknowledgements}
\begin{acknowledgements}

This research has made use of the Keck Observatory Archive (KOA), which is operated by the W. M. Keck Observatory and the NASA Exoplanet Science Institute (NExScI), under contract with the National Aeronautics and Space Administration. The data presented herein were obtained at the W.M. Keck Observatory, which is operated as a scientific partnership among the California Institute of Technology, the University of California and the National Aeronautics and Space Administration. The Observatory was made possible by the generous financial support of the W.M. Keck Foundation.

The authors wish to recognize and acknowledge the very significant cultural role and reverence that the summit of Maunakea has always had within the indigenous Hawaiian community. We are most fortunate to have the opportunity to conduct observations from this mountain. The authors wish to recognize and sincerely appreciate the work of the entire WMKO staff over the last two plus decades, and to the efforts of the team at the NASA Exoplanet Science Institute (NExScI) who are responsible for maintaining the KOA.

\end{acknowledgements}

\software{Astropy \citep{astropy2018}, Astroquery \citep{Ginsburg2019},
  Matplotlib \citep{hunter2007}, NumPy \citep{Harris2020}, pyransac \citep{Chum2003}, SciPy \citep{Virtanen2020}}

\clearpage

%%%%%%%%%%%%%%%%%%%%%%
%% Appendices

\appendix

In this Appendix, we provide information on \autoslit{}'s \slitastrometry\ routine in Section~\ref{sec:Details of slit astrometry Routine}. Section~\ref{sec:codeexample} outlines an example use of \tilsotua\ to reconstruct the sky position of a LRIS slitmask. The archived mask information is stored in FITS files with a similar structure to the equivalent DEIMOS and MOSFIRE files. In Section~\ref{sec:maskfiletable}, Table~\ref{tab:fitsextensiondetails} summarizes the information contained in each mask design FITS file extension.

\section{Details of \slitastrometry{} Routine}
\label{sec:Details of slit astrometry Routine}

The correction for focal plane distortions applied by \autoslit\ in the \slitastrometry{} subroutine is shown in equations \ref{eq:XCCD}-\ref{eq:YOUT} (Cohen and Shopbell 1996). First, \autoslit\ converts the X position in the mask frame to the X position in CCD pixels, $X_{CCD}$,

\begin{equation}
\label{eq:XCCD}
\begin{aligned}
X_{\rm CCD} &= N_{\rm pix} / 2.0 + (X_{\rm 0}^{\rm Mask} - X_{\rm Mask}^{\rm Final}) / s_{\rm CCD}.
\end{aligned}
\end{equation}

\noindent Here $s_{\rm CCD}$ is the scale of mm/pixel at the CCD and $N_{pix}$ is the number of pixels per side of the CCD. The focal plane distortion correction is then made using a second degree polynomial to obtain the corrected X position on the CCD, $X_{\rm CCD}^{\rm OUT}$,

\begin{equation}
\label{eq:XCCDout}
\begin{aligned}
X_{\rm CCD}^{\rm OUT} &= A_{\rm 1} + A_{\rm 2} X_{\rm CCD} + A_{\rm 3} Y_{\rm CCD} + A_{\rm 4} X_{\rm CCD}^{\rm 2} + A_{\rm 5}  Y_{\rm CCD} ^{\rm 2} + A_{\rm 6}  X_{\rm CCD} Y_{\rm CCD}.
\end{aligned}
\end{equation}

\noindent The A coefficients come from the \cite{cohen07} report, which first maps and then fits the distortion at the LRIS focal plane. The distortion corrected position in CCD pixels is then converted back to the mask coordinate system, $X_{\rm Mask}^{\rm Uncorr}$:

\begin{equation}
\label{eq:XOUT}
\begin{aligned}
X_{\rm Mask}^{\rm Uncorr} &= X_{\rm 0}^{\rm Mask} - (X_{\rm CCD}^{\rm OUT} - N_{\rm pix} / 2.0) s_{\rm CCD}.\\
\end{aligned}
\end{equation}

Similar steps are completed for the Y coordinate. The Y position on the mask, $Y_{\rm Mask}^{\rm Final}$, is converted to the raw Y position on the CCD, $Y_{\rm CCD}$ with

\begin{equation}
\label{eq:YCCD}
\begin{aligned}
Y_{\rm CCD} &= N_{\rm pix} / 2.0 - (Y_{\rm 0}^{\rm Mask} - Y_{\rm Mask}^{\rm Final}) / s_{\rm CCD}.\\
\end{aligned}
\end{equation}

\noindent The distortion corrected Y CCD position on the mask, $Y_{\rm CCD}^{\rm OUT}$ is calculated based on the \cite{cohen07} results, using 

\begin{equation}
\label{eq:YCCDout}
\begin{aligned}
Y_{\rm CCD}^{\rm OUT} &= B_{\rm 1} + B_{\rm 2} X_{\rm CCD} + B_{\rm 3}Y_{\rm CCD}.\\
\end{aligned}
\end{equation}

\noindent The B coefficients are taken from the focal plane distortion correction fit for the Y coordinate. Finally, \autoslit\ calculates the distortion corrected position in CCD pixels in the Y coordinate, $Y_{\rm Mask}^{\rm Uncorr}$:

\begin{equation}
\label{eq:YOUT}
\begin{aligned}
Y_{\rm Mask}^{\rm Uncorr} &= Y_{\rm 0}^{\rm Mask} + (Y_{\rm CCD}^{\rm OUT} - N_{\rm pix} / 2.0) s_{\rm CCD}.\\
\end{aligned}
\end{equation}

\tilsotua\ uses these equations to calculate the value of the correction made at the given point on the mask, $(X_{\rm Mask}^{\rm Uncorr},Y_{\rm Mask}^{\rm Uncorr})$ as described in \S~\ref{subsec:distortioncorrection}.

\section{\tilsotua\ Example}
\label{sec:codeexample}

The standard useage of \tilsotua\ is generating the LRIS slitmask sky position with the {\bf xytowcs} function. {\bf xytowcs} calls the subroutines handling the calculations described in Sections~\ref{subsec:xycoords}--\ref{sec:shifts} and generates the output files detailed in \ref{subsec:codeoutputs}. The following code block is an example script executing \tilsotua{}.

\includegraphics[width=\textwidth]{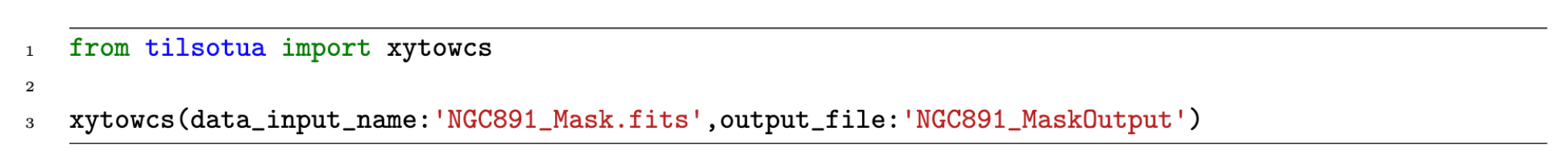}

{\bf xytowcs}'s first input is the UCO/Lick archival mask design file or the {\bf .file3} \autoslit\ output to provide the mask information. In this example, NGC891\_Mask.fits represents an archival mask design file from UCO/Lick. The second input, output\_file, serves as the base name of the four output files produced by \tilsotua{}.

Users can populate the target object information by providing {\bf .file1} \autoslit\ output and the catalog of objects provided during mask creation alongside the {\bf .file3} \autoslit\ output:

\includegraphics[width=\textwidth]{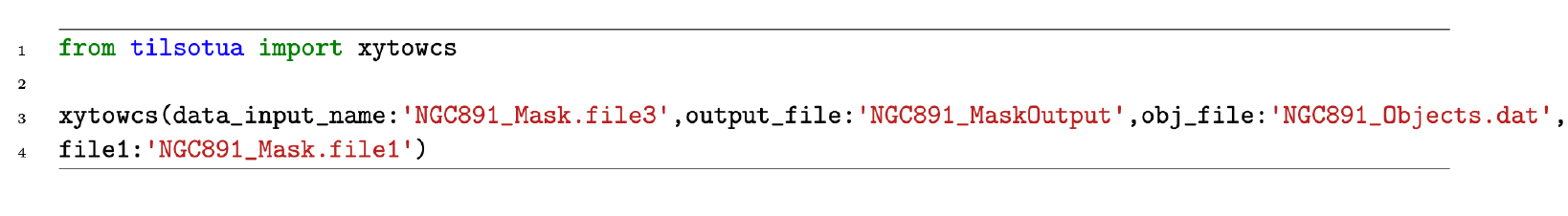}

In this example, NGC891\_Mask.file1 and NGC891\_Mask.file3 are the two \autoslit\ output files. NGC891\_Objects.dat contains the list of targets objects.

%%%%%%%%%%%%%%%%%
%% Appendix Tables

\clearpage

\section{Mask FITS File Extension Overview}
\label{sec:maskfiletable}
The archival LRIS mask FITS files are structured with multiple extensions containing the mask design information. Table~\ref{tab:fitsextensiondetails} contains the details of the information stored in each extension with a brief description of selected important keywords. \tilsotua\ output FITS files also follow this structure.

\begin{longdeluxetable}{lc}
\tablecaption{Mask FITS File Extension Details\label{tab:fitsextensiondetails}}
\tablehead{\colhead{Entry} & \colhead{Entry Description} }
\startdata
\multicolumn2c{Mask Design} \\
\hline
DesId & Mask Design ID Number \\
DesName & Mask Design Name \\
DesPId & \nodata\\
DesCreat & Design \autoslit\ Version \\
DesDate  & Mask Design Date\\
DesNslit & Number of Slits \\
DesNobj & \nodata\\
ProjName & Project Name\\
INSTRUME & Telescope Instrument (LRIS)\\
MaskType & \nodata\\
RA$\_$PNT & $\alpha_0$ of the Mask\\
Dec$\_$PNT & $\delta_0$ of the Mask\\
RADEPNT & Mask Coordinate Frame\\
EQUINPNT & Mask Coordinate Equinox\\
PA$\_$PNT & Mask Position Angle\\
DATE$\_$PNT & \nodata\\
LST$\_$PNT & \nodata\\
\hline
\multicolumn2c{Mask Owner} \\
\hline
obid & \nodata\\
firstnm & PI First Name\\
lastnm & PI Last Name\\
addr1/2 + City, State & PI Address\\
phone & PI Phone Number\\
email & PI Email\\
institution & PI Institution\\
comments &\nodata\\
inst & \nodata\\
privbits & \nodata\\
\hline
\multicolumn2c{DesiSlits} \\
\hline
dSlitId & Slit Design ID\\
DesId & Mask Design ID\\
slitRA & Slit $\alpha$\\
slitDec & Slit $\delta$\\
slitLen & Slit Length\\
slitLPA & Slit Length Position Angle \\
slitWid & Slit Width\\
slitWPA & Slit Width Position Angle\\
SlitName & Slit Name\\
\hline
\multicolumn2c{BluePrint} \\
\hline
BluId & Mask Blueprint ID\\
DesId & Mask Design ID\\
BluName & Mask Blueprint Name\\
BluPId & \nodata\\
BluCreat & Design Autoslit Version\\
BluDate & Blueprint Design Date\\
LST$\_$Use & \nodata\\
Date$\_$Use & Date Mask Used\\
TeleId & \nodata\\
RefrId &\nodata\\
AtmPres & Design Atmospheric Pressure\\
AtmHumid & Design Atmospheric Humidity\\
AtmTTLap &\nodata\\
RefWave& Design Reference Wavelength\\
DistId&\nodata\\
guiname&\nodata\\
millseq&\nodata\\
status&\nodata\\
loc&\nodata\\
\hline
\multicolumn2c{Mask Observer} \\
\hline
obid & \nodata\\
firstnm & Mask Observer First Name\\
lastnm & Mask Observer Last Name\\
addr1/2 + City, State & Mask Observer Address\\
phone & Mask Observer Phone Number\\
email & Mask Observer Email\\
institution & Mask Observer Institution\\
comments &\nodata\\
inst & \nodata\\
privbits & \nodata\\
\hline
\multicolumn2c{BluSlits} \\
\hline
bSlitId & Slit Blueprint ID\\
BluId & Mask Blueprint ID\\
dSlitId & Slit Design ID\\
slitX1 & $X_{\rm Mill}$ Position of First Slit Corner\\
slitY1 & $Y_{\rm Mill}$ Position of First Slit Corner\\
slitX2 & $X_{\rm Mill}$ Position of Second Slit Corner\\
slitY2 & $Y_{\rm Mill}$ Position of Second Slit Corner\\
slitX3 & $X_{\rm Mill}$ Position of Third Slit Corner\\
slitY3 & $Y_{\rm Mill}$ Position of Third Slit Corner\\
slitX4 & $X_{\rm Mill}$ Position of Fourth Slit Corner\\
slitY4 & $Y_{\rm Mill}$ Position of Fourth Slit Corner\\
bad & Flag\\
\enddata
\end{longdeluxetable}

\clearpage
\bibliography{references}

\begin{thebibliography}{}
\expandafter\ifx\csname natexlab\endcsname\relax\def\natexlab#1{#1}\fi
\providecommand{\url}[1]{\href{#1}{#1}}
\providecommand{\dodoi}[1]{doi:~\href{http://doi.org/#1}{\nolinkurl{#1}}}
\providecommand{\doeprint}[1]{\href{http://ascl.net/#1}{\nolinkurl{http://ascl.net/#1}}}
\providecommand{\doarXiv}[1]{\href{https://arxiv.org/abs/#1}{\nolinkurl{https://arxiv.org/abs/#1}}}

\bibitem[{{Bosler} {et~al.}(2007){Bosler}, {Smecker-Hane}, \&
  {Stetson}}]{bosler07}
{Bosler}, T.~L., {Smecker-Hane}, T.~A., \& {Stetson}, P.~B. 2007, \mnras, 378,
  318, \dodoi{10.1111/j.1365-2966.2007.11792.x}

\bibitem[{{Casey} {et~al.}(2008){Casey}, {Impey}, {Petry}, {Marble}, \&
  {Dav{\'e}}}]{casey08}
{Casey}, C.~M., {Impey}, C.~D., {Petry}, C.~E., {Marble}, A.~R., \& {Dav{\'e}},
  R. 2008, \aj, 136, 181, \dodoi{10.1088/0004-6256/136/1/181}

\bibitem[{{Chen} {et~al.}(2020){Chen}, {Steidel}, {Hummels}, {Rudie}, {Dong},
  {Trainor}, {Bogosavljevi{\'c}}, {Erb}, {Pettini}, {Reddy}, {Shapley},
  {Strom}, {Theios}, {Faucher-Gigu{\`e}re}, {Hopkins}, \&
  {Kere{\v{s}}}}]{chen20}
{Chen}, Y., {Steidel}, C.~C., {Hummels}, C.~B., {et~al.} 2020, \mnras, 499,
  1721, \dodoi{10.1093/mnras/staa2808}

\bibitem[{Chum {et~al.}(2003)Chum, Matas, \& Kittler}]{Chum2003}
Chum, O., Matas, J., \& Kittler, J. 2003, in Pattern Recognition

\bibitem[{{Cohen} \& {Huang}(2007)}]{cohen07}
{Cohen}, J.~D., \& {Huang}, W. 2007

\bibitem[{{Cohen} \& {Shopbell}(1996)}]{cohen96}
{Cohen}, J.~D., \& {Shopbell}, P. 1996

\bibitem[{{Cohen} {et~al.}(2000){Cohen}, {Hogg}, {Blandford}, {Cowie}, {Hu},
  {Songaila}, {Shopbell}, \& {Richberg}}]{cohen00}
{Cohen}, J.~G., {Hogg}, D.~W., {Blandford}, R., {et~al.} 2000, \apj, 538, 29,
  \dodoi{10.1086/309096}

\bibitem[{{Coil} {et~al.}(2011){Coil}, {Weiner}, {Holz}, {Cooper}, {Yan}, \&
  {Aird}}]{coil11}
{Coil}, A.~L., {Weiner}, B.~J., {Holz}, D.~E., {et~al.} 2011, \apj, 743, 46,
  \dodoi{10.1088/0004-637X/743/1/46}

\bibitem[{Collaboration {et~al.}(2018)Collaboration, Price-Whelan, Sip{\H o}cz,
  G{\"u}nther, Lim, Crawford, Conseil, Shupe, Craig, Dencheva, Ginsburg,
  VanderPlas, Bradley, P{\'e}rez-Su{\'a}rez, de~Val-Borro, Contributors),
  Aldcroft, Cruz, Robitaille, Tollerud, Committee), Ardelean, Babej, Bach,
  Bachetti, Bakanov, Bamford, Barentsen, Barmby, Baumbach, Berry, Biscani,
  Boquien, Bostroem, Bouma, Brammer, Bray, Breytenbach, Buddelmeijer, Burke,
  Calderone, Rodr{\'\i}guez, Cara, Cardoso, Cheedella, Copin, Corrales,
  Crichton, D'Avella, Deil, Depagne, Dietrich, Donath, Droettboom, Earl, Erben,
  Fabbro, Ferreira, Finethy, Fox, Garrison, Gibbons, Goldstein, Gommers, Greco,
  Greenfield, Groener, Grollier, Hagen, Hirst, Homeier, Horton, Hosseinzadeh,
  Hu, Hunkeler, Ivezi{\'c}, Jain, Jenness, Kanarek, Kendrew, Kern, Kerzendorf,
  Khvalko, King, Kirkby, Kulkarni, Kumar, Lee, Lenz, Littlefair, Ma, Macleod,
  Mastropietro, McCully, Montagnac, Morris, Mueller, Mumford, Muna, Murphy,
  Nelson, Nguyen, Ninan, N{\"o}the, Ogaz, Oh, Parejko, Parley, Pascual, Patil,
  Patil, Plunkett, Prochaska, Rastogi, Janga, Sabater, Sakurikar, Seifert,
  Sherbert, Sherwood-Taylor, Shih, Sick, Silbiger, Singanamalla, Singer,
  Sladen, Sooley, Sornarajah, Streicher, Teuben, Thomas, Tremblay, Turner,
  Terr{\'o}n, van Kerkwijk, de~la Vega, Watkins, Weaver, Whitmore, Woillez,
  Zabalza, \& Contributors)}]{astropy2018}
Collaboration, T.~A., Price-Whelan, A.~M., Sip{\H o}cz, B.~M., {et~al.} 2018,
  The Astronomical Journal, 156, 123.
\newblock \url{http://stacks.iop.org/1538-3881/156/i=3/a=123}

\bibitem[{{Dawson} {et~al.}(2004){Dawson}, {Rhoads}, {Malhotra}, {Stern},
  {Dey}, {Spinrad}, {Jannuzi}, {Wang}, \& {Landes}}]{dawson04}
{Dawson}, S., {Rhoads}, J.~E., {Malhotra}, S., {et~al.} 2004, \apj, 617, 707,
  \dodoi{10.1086/425572}

\bibitem[{{Faber} {et~al.}(2003){Faber}, {Phillips}, {Kibrick}, {Alcott},
  {Allen}, {Burrous}, {Cantrall}, {Clarke}, {Coil}, {Cowley}, {Davis}, {Deich},
  {Dietsch}, {Gilmore}, {Harper}, {Hilyard}, {Lewis}, {McVeigh}, {Newman},
  {Osborne}, {Schiavon}, {Stover}, {Tucker}, {Wallace}, {Wei}, {Wirth}, \&
  {Wright}}]{Faber2003}
{Faber}, S.~M., {Phillips}, A.~C., {Kibrick}, R.~I., {et~al.} 2003, in Society
  of Photo-Optical Instrumentation Engineers (SPIE) Conference Series, Vol.
  4841, Instrument Design and Performance for Optical/Infrared Ground-based
  Telescopes, ed. M.~{Iye} \& A.~F.~M. {Moorwood}, 1657--1669,
  \dodoi{10.1117/12.460346}

\bibitem[{{Flewelling} {et~al.}(2020){Flewelling}, {Magnier}, {Chambers},
  {Heasley}, {Holmberg}, {Huber}, {Sweeney}, {Waters}, {Calamida}, {Casertano},
  {Chen}, {Farrow}, {Hasinger}, {Henderson}, {Long}, {Metcalfe}, {Narayan},
  {Nieto-Santisteban}, {Norberg}, {Rest}, {Saglia}, {Szalay}, {Thakar},
  {Tonry}, {Valenti}, {Werner}, {White}, {Denneau}, {Draper}, {Hodapp},
  {Jedicke}, {Kaiser}, {Kudritzki}, {Price}, {Wainscoat}, {Chastel}, {McLean},
  {Postman}, \& {Shiao}}]{flewelling20}
{Flewelling}, H.~A., {Magnier}, E.~A., {Chambers}, K.~C., {et~al.} 2020, \apjs,
  251, 7, \dodoi{10.3847/1538-4365/abb82d}

\bibitem[{{Gaia Collaboration} {et~al.}(2016){Gaia Collaboration}, {Prusti},
  {de Bruijne}, {Brown}, {Vallenari}, {Babusiaux}, {Bailer-Jones}, {Bastian},
  {Biermann}, {Evans}, \& et~al.}]{GAIACollaboration2016b}
{Gaia Collaboration}, {Prusti}, T., {de Bruijne}, J.~H.~J., {et~al.} 2016,
  \aap, 595, A1, \dodoi{10.1051/0004-6361/201629272}

\bibitem[{{Gaia Collaboration} {et~al.}(2023){Gaia Collaboration}, {Vallenari},
  {Brown}, {Prusti}, {de Bruijne}, {Arenou}, {Babusiaux}, {Biermann},
  {Creevey}, {Ducourant}, \& et~al.}]{GAIACollaboration2023j}
{Gaia Collaboration}, {Vallenari}, A., {Brown}, A.~G.~A., {et~al.} 2023, \aap,
  674, A1, \dodoi{10.1051/0004-6361/202243940}

\bibitem[{{Ginsburg} {et~al.}(2019){Ginsburg}, {Sip{\H{o}}cz}, {Brasseur},
  {Cowperthwaite}, {Craig}, {Deil}, {Guillochon}, {Guzman}, {Liedtke}, {Lian
  Lim}, {Lockhart}, {Mommert}, {Morris}, {Norman}, {Parikh}, {Persson},
  {Robitaille}, {Segovia}, {Singer}, {Tollerud}, {de Val-Borro}, {Valtchanov},
  {Woillez}, {Astroquery Collaboration}, \& {a subset of astropy
  Collaboration}}]{Ginsburg2019}
{Ginsburg}, A., {Sip{\H{o}}cz}, B.~M., {Brasseur}, C.~E., {et~al.} 2019, \aj,
  157, 98, \dodoi{10.3847/1538-3881/aafc33}

\bibitem[{{Harish} {et~al.}(2021){Harish}, {Wold}, {Rhoads}, {Malhotra}, {Hu},
  {Wang}, {Zheng}, {Infante}, {Barrientos}, {Walker}, \& {Lager
  Team}}]{Harish2021}
{Harish}, S., {Wold}, I., {Rhoads}, J., {et~al.} 2021, in American Astronomical
  Society Meeting Abstracts, Vol. 238, American Astronomical Society Meeting
  Abstracts, 331.07

\bibitem[{{Harris} {et~al.}(2020){Harris}, {Millman}, {van der Walt},
  {Gommers}, {Virtanen}, {Cournapeau}, {Wieser}, {Taylor}, {Berg}, {Smith},
  {Kern}, {Picus}, {Hoyer}, {van Kerkwijk}, {Brett}, {Haldane}, {del R{\'\i}o},
  {Wiebe}, {Peterson}, {G{\'e}rard-Marchant}, {Sheppard}, {Reddy}, {Weckesser},
  {Abbasi}, {Gohlke}, \& {Oliphant}}]{Harris2020}
{Harris}, C.~R., {Millman}, K.~J., {van der Walt}, S.~J., {et~al.} 2020, \nat,
  585, 357, \dodoi{10.1038/s41586-020-2649-2}

\bibitem[{{Hu} {et~al.}(2004){Hu}, {Cowie}, {Capak}, {McMahon}, {Hayashino}, \&
  {Komiyama}}]{hu04}
{Hu}, E.~M., {Cowie}, L.~L., {Capak}, P., {et~al.} 2004, \aj, 127, 563,
  \dodoi{10.1086/381302}

\bibitem[{{Hunter}(2007)}]{hunter2007}
{Hunter}, J.~D. 2007, Computing in Science and Engineering, 9, 90,
  \dodoi{10.1109/MCSE.2007.55}

\bibitem[{{Kalirai} {et~al.}(2007){Kalirai}, {Bergeron}, {Hansen}, {Kelson},
  {Reitzel}, {Rich}, \& {Richer}}]{Kalirai07}
{Kalirai}, J.~S., {Bergeron}, P., {Hansen}, B. M.~S., {et~al.} 2007, \apj, 671,
  748, \dodoi{10.1086/521922}

\bibitem[{{Lee} {et~al.}(2018){Lee}, {Krolewski}, {White}, {Schlegel},
  {Nugent}, {Hennawi}, {M{\"u}ller}, {Pan}, {Prochaska}, {Font-Ribera},
  {Suzuki}, {Glazebrook}, {Kacprzak}, {Kartaltepe}, {Koekemoer}, {Le
  F{\`e}vre}, {Lemaux}, {Maier}, {Nanayakkara}, {Rich}, {Sanders}, {Salvato},
  {Tasca}, \& {Tran}}]{lee18}
{Lee}, K.-G., {Krolewski}, A., {White}, M., {et~al.} 2018, \apjs, 237, 31,
  \dodoi{10.3847/1538-4365/aace58}

\bibitem[{{Martin} \& {Bouch{\'e}}(2009)}]{martin09}
{Martin}, C.~L., \& {Bouch{\'e}}, N. 2009, \apj, 703, 1394,
  \dodoi{10.1088/0004-637X/703/2/1394}

\bibitem[{{Mazzucchelli} {et~al.}(2017){Mazzucchelli}, {Ba{\~n}ados},
  {Venemans}, {Decarli}, {Farina}, {Walter}, {Eilers}, {Rix}, {Simcoe},
  {Stern}, {Fan}, {Schlafly}, {De Rosa}, {Hennawi}, {Chambers}, {Greiner},
  {Burgett}, {Draper}, {Kaiser}, {Kudritzki}, {Magnier}, {Metcalfe}, {Waters},
  \& {Wainscoat}}]{Mazzucchelli2017}
{Mazzucchelli}, C., {Ba{\~n}ados}, E., {Venemans}, B.~P., {et~al.} 2017, \apj,
  849, 91, \dodoi{10.3847/1538-4357/aa9185}

\bibitem[{{McCarthy} {et~al.}(1998){McCarthy}, {Cohen}, {Butcher}, {Cromer},
  {Croner}, {Douglas}, {Goeden}, {Grewal}, {Lu}, {Petrie}, {Weng}, {Weber},
  {Koch}, \& {Rodgers}}]{mccarthy1998}
{McCarthy}, J.~K., {Cohen}, J.~G., {Butcher}, B., {et~al.} 1998, in Society of
  Photo-Optical Instrumentation Engineers (SPIE) Conference Series, Vol. 3355,
  Optical Astronomical Instrumentation, ed. S.~{D'Odorico}, 81--92,
  \dodoi{10.1117/12.316831}

\bibitem[{{Oke} {et~al.}(1995){Oke}, {Cohen}, {Carr}, {Cromer}, {Dingizian},
  {Harris}, {Labrecque}, {Lucinio}, {Schaal}, {Epps}, \& {Miller}}]{oke1995}
{Oke}, J.~B., {Cohen}, J.~G., {Carr}, M., {et~al.} 1995, \pasp, 107, 375,
  \dodoi{10.1086/133562}

\bibitem[{{Prochaska} {et~al.}(2020{\natexlab{a}}){Prochaska}, {Hennawi},
  {Westfall}, {Cooke}, {Wang}, {Hsyu}, {Davies}, {Farina}, \&
  {Pelliccia}}]{Prochaska2020a}
{Prochaska}, J., {Hennawi}, J., {Westfall}, K., {et~al.} 2020{\natexlab{a}},
  The Journal of Open Source Software, 5, 2308, \dodoi{10.21105/joss.02308}

\bibitem[{{Prochaska} {et~al.}(2020{\natexlab{b}}){Prochaska}, {Hennawi},
  {Cooke}, {Westfall}, {Wang}, {EmAstro}, {Tiffanyhsyu}, {Wasserman},
  {Villaume}, {Marijana777}, {Schindler}, {Young}, {Simha}, {Wilde}, {Tejos},
  {Isbell}, {Fl{\"o}rs}, {Sandford}, {Vasovi{\'c}}, {Betts}, \&
  {Holden}}]{pypeit:zenodo}
{Prochaska}, J.~X., {Hennawi}, J., {Cooke}, R., {et~al.} 2020{\natexlab{b}},
  {pypeit/PypeIt: Release 1.0.0}, v1.0.0,  Zenodo,
  \dodoi{10.5281/zenodo.3743493}

\bibitem[{{Reddy} {et~al.}(2006){Reddy}, {Steidel}, {Erb}, {Shapley}, \&
  {Pettini}}]{reddy06}
{Reddy}, N.~A., {Steidel}, C.~C., {Erb}, D.~K., {Shapley}, A.~E., \& {Pettini},
  M. 2006, \apj, 653, 1004, \dodoi{10.1086/508851}

\bibitem[{{Reitzel} {et~al.}(2004){Reitzel}, {Guhathakurta}, \&
  {Rich}}]{reitzel04}
{Reitzel}, D.~B., {Guhathakurta}, P., \& {Rich}, R.~M. 2004, \aj, 127, 2133,
  \dodoi{10.1086/382517}

\bibitem[{{Rubin} {et~al.}(2010){Rubin}, {Prochaska}, {Koo}, {Phillips}, \&
  {Weiner}}]{rubin10}
{Rubin}, K. H.~R., {Prochaska}, J.~X., {Koo}, D.~C., {Phillips}, A.~C., \&
  {Weiner}, B.~J. 2010, \apj, 712, 574, \dodoi{10.1088/0004-637X/712/1/574}

\bibitem[{{Rudie} {et~al.}(2019){Rudie}, {Steidel}, {Pettini}, {Trainor},
  {Strom}, {Hummels}, {Reddy}, \& {Shapley}}]{rudie19}
{Rudie}, G.~C., {Steidel}, C.~C., {Pettini}, M., {et~al.} 2019, \apj, 885, 61,
  \dodoi{10.3847/1538-4357/ab4255}

\bibitem[{{Shetrone} {et~al.}(2010){Shetrone}, {Martell}, {Wilkerson}, {Adams},
  {Siegel}, {Smith}, \& {Bond}}]{shetrone10}
{Shetrone}, M., {Martell}, S.~L., {Wilkerson}, R., {et~al.} 2010, \aj, 140,
  1119, \dodoi{10.1088/0004-6256/140/4/1119}

\bibitem[{{Shibuya} {et~al.}(2014){Shibuya}, {Ouchi}, {Nakajima}, {Hashimoto},
  {Ono}, {Rauch}, {Gauthier}, {Shimasaku}, {Goto}, {Mori}, \&
  {Umemura.}}]{Shibuya2014}
{Shibuya}, T., {Ouchi}, M., {Nakajima}, K., {et~al.} 2014, \apj, 788, 74,
  \dodoi{10.1088/0004-637X/788/1/74}

\bibitem[{{Steidel} {et~al.}(2011){Steidel}, {Bogosavljevi{\'c}}, {Shapley},
  {Kollmeier}, {Reddy}, {Erb}, \& {Pettini}}]{steidel11}
{Steidel}, C.~C., {Bogosavljevi{\'c}}, M., {Shapley}, A.~E., {et~al.} 2011,
  \apj, 736, 160, \dodoi{10.1088/0004-637X/736/2/160}

\bibitem[{{Steidel} {et~al.}(2010){Steidel}, {Erb}, {Shapley}, {Pettini},
  {Reddy}, {Bogosavljevi{\'c}}, {Rudie}, \& {Rakic}}]{steidel10}
{Steidel}, C.~C., {Erb}, D.~K., {Shapley}, A.~E., {et~al.} 2010, \apj, 717,
  289, \dodoi{10.1088/0004-637X/717/1/289}

\bibitem[{{Steidel} {et~al.}(2004){Steidel}, {Shapley}, {Pettini},
  {Adelberger}, {Erb}, {Reddy}, \& {Hunt}}]{steidel2004}
{Steidel}, C.~C., {Shapley}, A.~E., {Pettini}, M., {et~al.} 2004, \apj, 604,
  534, \dodoi{10.1086/381960}

\bibitem[{{Virtanen} {et~al.}(2020){Virtanen}, {Gommers}, {Oliphant},
  {Haberland}, {Reddy}, {Cournapeau}, {Burovski}, {Peterson}, {Weckesser},
  {Bright}, {van der Walt}, {Brett}, {Wilson}, {Millman}, {Mayorov}, {Nelson},
  {Jones}, {Kern}, {Larson}, {Carey}, {Polat}, {Feng}, {Moore}, {VanderPlas},
  {Laxalde}, {Perktold}, {Cimrman}, {Henriksen}, {Quintero}, {Harris},
  {Archibald}, {Ribeiro}, {Pedregosa}, {van Mulbregt}, \& {SciPy 1. 0
  Contributors}}]{Virtanen2020}
{Virtanen}, P., {Gommers}, R., {Oliphant}, T.~E., {et~al.} 2020, Nature
  Methods, 17, 261, \dodoi{10.1038/s41592-019-0686-2}

\bibitem[{{Werk} {et~al.}(2012){Werk}, {Prochaska}, {Thom}, {Tumlinson},
  {Tripp}, {O'Meara}, \& {Meiring}}]{werk12}
{Werk}, J.~K., {Prochaska}, J.~X., {Thom}, C., {et~al.} 2012, \apjs, 198, 3,
  \dodoi{10.1088/0067-0049/198/1/3}

\end{thebibliography}

%%%%%%%%%%%%%%%%%%%%%%%%%%%%%%%%%%%%%%%%%%%%%%%%%%%%%%%%%%%%%%%%%%%%%%

\clearpage

\end{document}